\documentclass{ieee}
\usepackage[T1]{fontenc}
\usepackage{mathptmx}
\usepackage[scaled=0.92]{helvet}
\usepackage{hyperref,url}
\usepackage[dvips]{graphicx}
\usepackage{subfigure}
\usepackage{bm}
\newcommand{\la}{\lambda}
\newcommand{\La}{\bm \Lambda}

\begin{document}
\date{}
%
\title{\Large\bf Mapping the Physical Properties of Cosmic Hot Gas with Hyper-spectral Imaging }
%
%
\author{\begin{tabular}[t]{c@{\extracolsep{8em}}c} 
Mark O'Dwyer  & Trevor Ponman \\
Ela Claridge & Somak Raychaudhury
 \\
        School of Computer Science & School of Physics and Astronomy\\
        The University of Birmingham & The University of Birmingham\\
        Birmingham B15 2TT, U.K. & Birmingham B15 2TT, U.K. 
\end{tabular}}

\maketitle

\subsection*{\centering  Abstract}

{\em
A novel  inversion technique is proposed to compute parametric maps  showing the temperature, density and chemical composition of cosmic hot gas from X-ray hyper-spectral images. The parameters are recovered by constructing a unique non-linear mapping derived by combining a physics-based modelling of the X-ray spectrum with the selection of optimal bandpass filters. Preliminary results and analysis are presented.
}

\section{Introduction}

With the advent of space-borne X-ray astronomy in the 1960s, it became apparent that there are many cosmic sources emitting high energy electromagnetic radiation, indicating the presence of matter at very high temperatures. Point-like X-ray sources arise from hot gas falling onto collapsed objects: white dwarfs, neutron stars and black holes, but sources of diffuse X-ray emission were also discovered. Such extended sources generally indicate the presence of clouds of gas at temperatures above $10^6$ K. Examples (on increasing size scales) include (a) supernova remnants, resulting from the explosion of massive dying stars, (b) hot gas within galaxies, often heated by supernova explosions, and (c) gas trapped in the deep gravitational potential wells of clusters of galaxies, typically 10 million light years in size, which are the largest stable structures in the Universe. We will deal with observations of (c) in this work.

X-ray telescopes typically consist of a mirror system, focusing X-rays onto a detector, which records the position and energy of each detected photon. The basic data product, when such a telescope is used to observe an extended cosmic X-ray source, is a spectral map of the emission from the hot gas in the target. The spectral properties of this emission provide information about the temperature of the gas, and also about its chemical composition. In principle, therefore, information is available which would enable the temperature and composition of the gas to be mapped, enabling astronomers to investigate the origin of the gas, and the mechanisms which have heated it.

Unfortunately, unlike optical light for cosmic sources, X-ray photons are difficult to collect, since the earth's atmosphere absorbs and scatters them and thus X-ray observatories have to be space-borne.  Furthermore, X-ray photons are scarce even from the brightest sources, and a long observation of a comparatively bright source with a major X-ray observatory may detect only tens of thousands of photons. The issue of extracting reliable spatial-spectral information under these circumstances is not straightforward, and the methods commonly employed by astronomers have tended to be very CPU-intensive, and to produce less than satisfactory results, as will be described below. 

With the advent of large publicly available astronomical archives (the
``Virtual Observatory", \url{http://www.ivoa.net}), it has become
necessary to produce algorithms for the fast and flexible extraction
of maps of diffuse emission. We are developing a multi-spectral
inversion approach which will allow such extraction of key gas
parameters, complete with statistical errors, once an initial
inversion mapping has been calculated. A similar approach could be
employed in other applications, wherever a well-defined model is
available, linking physical properties to the resulting spectra.

\begin{figure*}[htb!]
\begin{center}
\includegraphics[height=5.8cm,origin=lt,keepaspectratio]{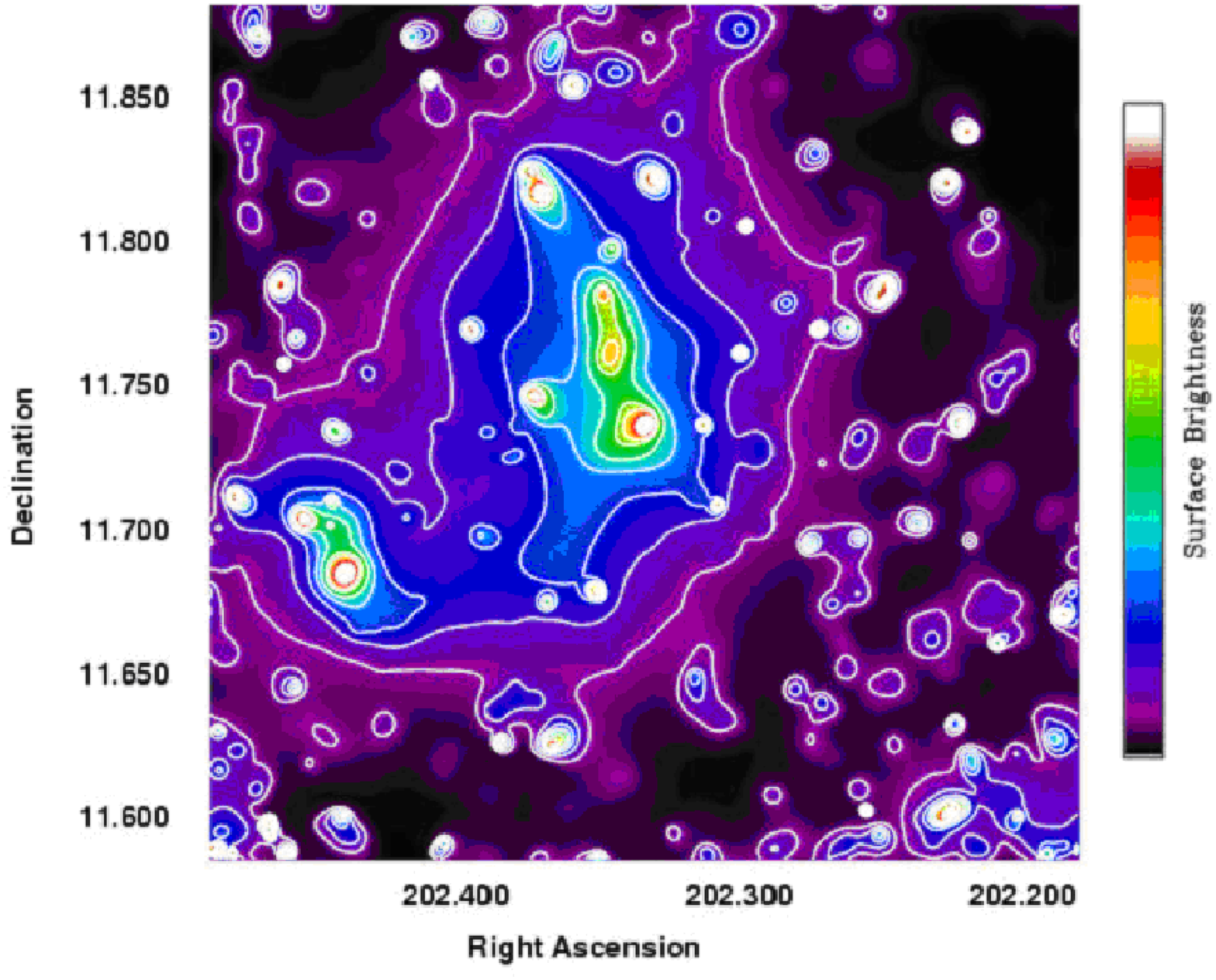}
\includegraphics[height=5.1cm,origin=lt,keepaspectratio]{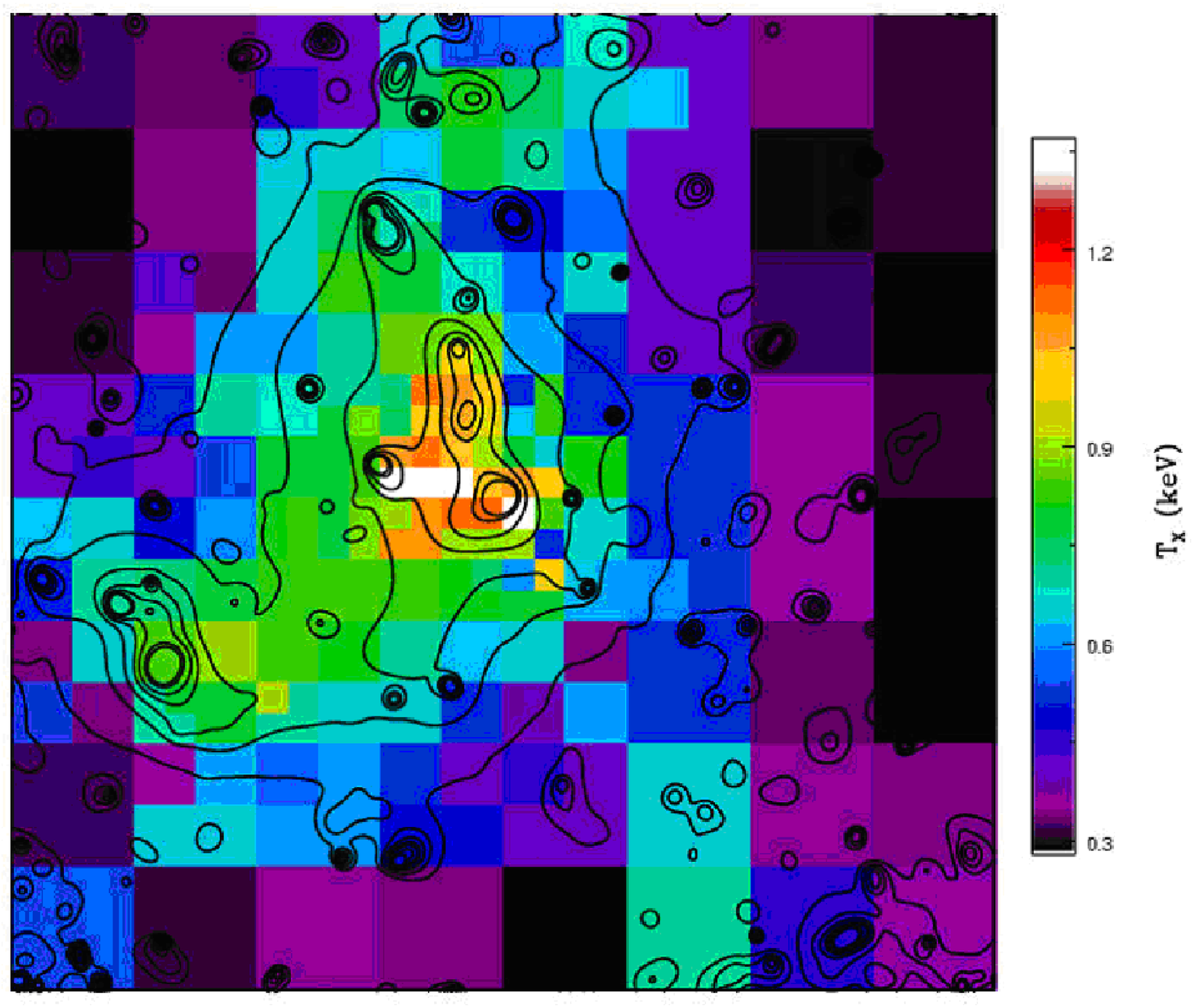}
%
%
%
\caption{\small (a) Adaptively smoothed 
X-ray image of the galaxy group 
centred on the galaxy NGC 5171, derived from data taken with the 
European Space Agency's XMM-Newton X-ray Observatory 
(http://xmm.vilspa.esa.es/); 
(b) Gas temperature map (white=hot to purple=cool) derived from 
the XMM-Newton observation of the galaxy group NGC5171 obtained  
by detailed spectral fitting. Contours show the surface brightness 
distribution from (a).}
\end{center}
\end{figure*}

\section{Physical mapping of hot gas in galaxy clusters}

The processes which generate X-ray emission in a hot plasma (a
partially ionised gas) are complex and varied. Extensive numerical
codes (e.g. [1]) are available to calculate the intricate web of
atomic excitations and de-excitations in the plasma, and compute the
emergent spectrum under circumstances where the plasma is ``optically
thin" (re-absorption of photons in the hot plasma is negligible) and
in thermodynamic equilibrium (i.e. has had sufficient time for the
populations of the different ionisation states to achieve stability).
Fortunately, these are good approximations in many astrophysical
contexts, due to the very low density of the gas, and the long
evolutionary timescales of the systems concerned.

\subsection{Interpreting a spectral map of intergalactic gas}

In this paper we consider the task of mapping the physical properties of the hot intergalactic gas within a small cluster (or ``group") of galaxies. An X-ray image of this gas is shown in Fig.~1(a) for the group of galaxies surrounding the nearby galaxy NGC 5171 [2]. The X-ray image shown in Fig.~1(a)  has been processed in order to suppress the ``Poisson noise", which arises from the limited number of photons available, using an adaptive smoothing technique. This involves convolving the raw image with a kernel whose width is small in regions of high brightness, but broad in regions of low brightness, where statistics are poor. The brightness of the emission is primarily an indication of the density of the hot gas, since the emissivity scales as the square of the gas density. 

In order to derive other physical parameters, it is necessary to extract spectra, and to compare these with one of the hot plasma models referred to above. These models are parameterised, in simple cases, by a small number of parameters - typically the gas temperature $T$, the abundance $Z$ of heavy elements relative to the standard fractions seen within the solar system, and a normalisation constant from which the projected density of the gas can be derived. Absorption of some X-rays en route to us, by cool gas within our own Milky Way, acts to preferentially reduce the intensity of low energy X-rays, in a way which is well understood, and this introduces a further parameter into the spectral model (corresponding to the projected density $n_H$ of cold gas along the line of sight to our target). Hence, in this simple case, the absorbed plasma model has four free parameters. For any choice of these parameters, the model can be used to predict a spectrum which is incident on the telescope, this is then folded through the spectral response of the instrument, to produce a prediction for the detected spectrum, which is compared with that actually observed (after the removal of any background emission). The parameters of the spectral model are then optimised in an iterative process, and errors on these parameters derived from the way in which the fit statistic deteriorates as parameters are moved away from their best fit values.

\subsection{Current mapping method}

In order to map the physical properties of the gas over the image, it is normally necessary to extract spectra from a set of spatial regions [2]. However a minimum number of detected photons (approx. 1000) are required in order to derive meaningful constraints from the spectral fitting process. One therefore has to extract spectra from regions of varying size (larger in areas of low brightness), fit each individually, and then assemble the results into an image in order to visualise the physical structure of the hot gas. This process is very time consuming, and produces a result which is very pixellated and hard to interpret (compare, for example, the blocky temperature map of Fig.~1(b) with the continuous brightness distribution of Fig.~1(a)).

\section{A new parameter recovery method}

In a new method of parameter recovery, a set of correspondences is established between the physical parameters and multi-band  image vectors. This is done in two steps. First, spectra are computed from parameters specifying object properties (see 2.1). Second, the equivalent image vectors are computed by convolving the spectra with suitably chosen bandpass filters. These two steps are computationally intensive, however they need to be carried out only once for a given imaging set-up. The set of correspondences established in this way forms the imaging model. The model is used to perform the inversion process, i.e. to infer the combination of parameters which lead to a particular image vector. This last step, which effectively performs image interpretation, is very fast and in some instances can be performed in real time. Its result is a set of parametric maps which show the magnitude of each parameter in a separate 2-dimensional image. In addition to improved speed, the ability to smooth the bandpass filtered images in a manner similar to that used in Fig.~1(a), will enable us to generate continuously varying parameter maps representing far better the continuous distributions expected in the hot gas.

\subsection{Modelling of the spectra}

Using widely available software [1] synthetic spectra were generated for X-ray emitting optically thin plasmas with the range of  physical parameters covering the range of interest, namely: the temperature $T$ (0.6--1.5 keV), elemental abundance $Z$ (0-0.9 times their corresponding solar values), and line-of-sight absorption due to a column of hydrogen gas $n_H$ (0--$1.8\times10^{25} m^{-2}$). Fig.~2 shows three examples illustrating how the shape and detailed structure of the spectra in the range 0.2--4 keV vary with various values of these parameters.  The fourth parameter, the overall normalisation, is adjusted to keep the total number of photons detected to a constant value. The spectra were then convolved with the telescope and detector response functions of the Advanced CCD Imaging spectrometer of the Chandra X-ray Observatory, the instrument used to obtain the observed data of HCG~62 used in the following section.

\begin{figure}[thb!] 
\begin{center} 
\includegraphics[width=0.95\hsize]{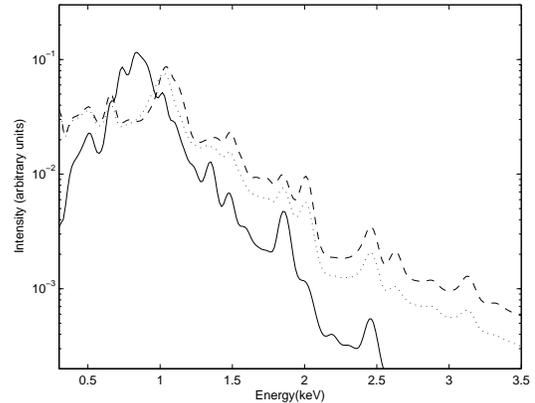} 
\caption{\small Three examples of the spectra used to construct our filters. They correspond to the following values for the three parameters ($T$ in keV, absorbing column  $n_H$ in $10^{25} m^{-2}$,  abundance $Z$ as fraction of solar)= full line (0.6, 1.0, 0.3); dashed line (1.5, 0.4, 0.6) and dotted line (1.2, 0.4, 0.3).}
\label{fig2} 
\end{center} 
\end{figure}

\subsection{Computational forward model}
	
Let an instance of hyper-spectral X-ray image data be represented by a discrete vector 
\begin{equation}
\la = \{\la_m\},   m=1,...,M, \qquad {\bf \la} \in {\bf \La} \\
\end{equation}
and the parameters depicting the physical properties be represented by a vector
\begin{equation}
{\bf p} = \{p_k\},  k=1,...,K, \qquad {\bf p} \in {\bf P} \\
\end{equation}
The mapping $f$
\begin{equation}
f: P \longrightarrow {\bf \La}
\end{equation}
models the physical processes governing the formation of spectra, for example the hot plasma model described in 2.

In the first step of the construction of the imaging model, $f$ is used to generate a set of spectra corresponding to the entire range of the physically valid instances of parameters ${\bf p}$, suitably discretised. Fig.~2 shows three examples of spectra generated using the model.

In the existing parameter recovery method (see 2.2) the parameter vector ${\bf p}$ is recovered directly from the hyper-spectral data by iteratively minimising the distance between the measured and the computed spectrum. Given the complexity of the model $f$ and the large dimensionality (M) of the spectra space ${\bf \La}$, this process is inevitably slow. Reducing M is likely to speed-up the process, but at the cost of loss of accuracy. However, provided certain conditions are fulfilled, it is possible to reduce the dimensionality in such a way that relatively accurate information regarding the original parameter values can still be recovered [4]. 

Mapping $g$
\begin{equation}
g: \La \longrightarrow I  \qquad I \in R^n \\
\end{equation}
defines a projection from M-dimensional X-ray spectra space ${\bf \La}$ into 
N-dimensional bandpass image space I such that N $\ll$ M. The mapping 
is implemented as bandpass filtering, through a convolution of the 
spectral vector 
$\la \in R^M$
with N filter response functions. 
The response function for the $n$-th filter can be defined as
\begin{equation}
i_n = \sum^{M}_{m=1} C^n_m \la_m, {\bf i} = \{i_n\}, n=1,...,N
\end{equation}
The mapping function $g$ can be defined by specifying the elements of 
matrix $C^n_m$ such as to meet the required optimisation criteria, as will be discussed in the next section.

With the introduction of $g$, the mapping
\begin{equation}
h = f \circ g  \qquad : P \longrightarrow I
\end{equation}
describes the correspondence between specific parameters characterising 
the astronomical object and the low-dimensional image vector 
computed by convolving the large-dimensional hyper-spectral image through 
a set of optimally chosen filters $C^n_m$. Whereas $f$ is determined by 
physics, and in this sense ``fixed", $g$ can be altered to meet user-defined optimisation criteria.

\subsection{Optimal spectral filter selection}

Drawing on the concepts introduced in 3.2 above, the task of estimating the parameters which characterise an object is to find an inverse function
\begin{equation}
h^{-1}: I \longrightarrow P
\end{equation}
such that $h^{-1}$ is unique, ${\bf p} = h^{-1}({\bf i})$ exists for all 
$\qquad {\bf p} \in P$ and the error of mapping from the image data to the parameter data is at its minimum. 

The uniqueness of mapping is the necessary condition and is tested by examining the behaviour of the determinant of the Jacobian matrix over all 
${\bf p} \in P$. One-to-one correspondence is ensured if and only if 
$h$ is monotonic on the entire domain P of {\bf p}. If $h$ were continuous, 
this can be easily determined by ensuring that all partial derivatives of 
$h$ are strictly larger than zero over P. When, as in this case, 
$h$ is a discrete vector valued function of a vector variable, the inverse 
function theorem [5] can be used to define an equivalent set of conditions: 
\begin{enumerate}
\item
$det(J) \neq 0  ~~for~all~~ \qquad {\bf p} \in P$,\\
where J = $\displaystyle{ \partial i_n \over \partial p_k}$, n=1,...,N; k=1,...,K, is a Jacobian matrix of discrete partial derivatives
\item
the sign of det(J) is the same over the whole domain of $\qquad {\bf p} \in P$ [4].
\end{enumerate}

If these conditions are fulfilled, then, according to [5], there exists 
a neighbourhood around each ${\bf p} \in P$ where there is one-to-one 
mapping between parameters {\bf p} and image vectors {\bf i}. Within such 
neighbourhood the inverse function $h^{-1}$ can be expressed as
\begin{equation}
d{\bf p} = J^{-1}d{\bf i}
\end{equation}
where  $d{\bf p} = {\bf p} - {\bf p}_0$,   $d{\bf i} = {\bf i} - {\bf i}_0$   
and  ${\bf i}_0 = h({\bf p}_0)$.

There could be many functions $h$ meeting the uniqueness criterion. 
However, due to the errors inherent in both the model and the measurements, 
the parameter recovery errors will be different for different choices of 
$h= f \circ g$. Since only the $g$ component can be manipulated (see 3.2), 
the overall effect of $h$ is determined by the choice of filter functions, 
$C^n_m$. The optimal mapping function is defined as the one which minimises the parameter recovery error. 

Using the appropriate statistical analysis [6] error in the k-th parameter 
can be estimated to be: \\
\begin{center}
$\epsilon_k = (( \displaystyle{ \partial p_k \over \partial i_n} )^2 (\epsilon^{poiss}_k)^2 )^{1 \over 2}$, 
$\epsilon^{poiss}_k = \sum^{M}_{m=1} C^n_m P_m$ 
\end{center}

$P_m$ is error corresponding to Poisson noise at energy level m, and 
$\displaystyle{ \partial p_k \over \partial i_n}$
are available from the inverse of the Jacobian matrix, $J^{-1}$. 
The overall error is computed as the sum of $\epsilon_k$.

The outline of the algorithm used for finding optimal filters is given below. The optimisation procedure was implemented using a technique based on Evolutionary Strategy [7], although other optimization algorithms could have been used with similar effect. Simple rectangular filters were used, each filter defined by its minimum and maximum energy. \\
\\
Until the stopping criterion is met\\
1.~define a new set of filters $C^n_m$\\
2.~for a given set of $C^n_m$ and
for each discrete ${\bf p}$
compute ${\bf i}$ = g($C^n_m$, f({\bf p}))\\
3.~check that $J$ is either strictly positive or strictly negative for ALL {\bf p} \\
if true, 
compute the inverse Jacobian matrix\\
if false, return to step 1\\
4.~compute the error of parameter recovery and use it to compute a stopping criterion.\\

The use of the Jacobian matrix requires that the number of filters (N) is 
equal to the number of parameters (K). One of the parameters, 
the normalisation constant, is required to compensate for the uneven 
energy levels across the image space due to the nature of the imaging 
process and properties of the imaging equipment. For this reason, 
instead of computing the image vector in step 2, an image quotient vector, 
defined as
$q_n$ = $\displaystyle{  i_n \over i_N}$,  n=1,...N-1
was computed and the Jacobian matrix was defined as 
\begin{equation}
J = {\partial q_n \over \partial p_k}, 
n=1,...,N-1; k=1,...,K; K = N-1
\end{equation}
\begin{figure}[thb!] 
\begin{center} 
\includegraphics[width=0.7\hsize]{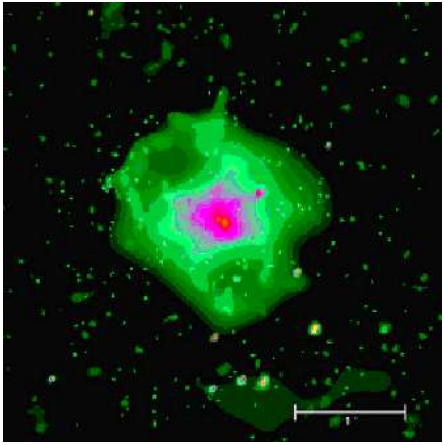} 
\caption{\small 
A hyper-spectral X-ray image of the compact galaxy group HCG~62, 
in the constellation Virgo. The image is 4 arcmin on a side.
Colours represent various levels of  X-ray surface brightness: green depicts lower brightness, while purple indicates increasing X-ray intensity.
\label{fig3} }
\end{center} 
\end{figure}

\subsection{Implementing mapping functions $h$ and $h^{-1}$}

The selection of optimal filters completes the modelling process. All 
the spectra generated with the model $f$ and corresponding to all the 
valid parameter vectors {\bf p} (see 3.1)  are now convolved with 
the filters (mapping $g$, see 3.3), the resulting image quotients {\bf q} 
are then computed and stored in a discrete look-up table indexed by {\bf p}. 
This table implements the (forward) mapping function $h$, such that 
{\bf q} = $h$({\bf p}) for all ${\bf p} \in P$. As a set of parameter 
vectors {\bf p} used to generate a specific embodiment of $h$ is discrete, 
the computation of $h$ for the parameters which are not in the generating 
set involves suitable interpolation (e.g. [8]).

The inverse mapping function $h^{-1} : I \longrightarrow P$
is implemented as 
a simple look-up, i.e. given a quotient vector 
${\bf q} =   \{\frac{i_n}{i_{N}}\}$  , n=1,...N-1,, find ${\bf p}_k$ which 
minimises the difference
$\| {\bf q}_n - $h$({\bf p}_k) \|$. 
As in the forward case, interpolation will produce a more accurate result.

\subsection{Creating parametric maps}
Parametric maps are computed as follows:\\
1.~Pre-process hyper-spectral images (compensate for read-out noise, remove instrument related artefacts etc.)\\
2.~Filter the images through $C^{N}_M$ to obtain N bandpass images\\
3.~Apply  spatial smoothing to the bandpass images\\
4.~Apply exposure correction\\
5.~Compute image quotients by dividing the images in the first N-1 bands by the image in $N^{th}$ band\\
6.~For each image location recover the parameter vector from the image quotient, {\bf p} = $h^{-1}({\bf q})$\\
7.~Store each component of the parameter vector ${\bf p}_k$, in a separate image (a parameteric map k)

\begin{figure}[thb!] 
\begin{center} 
\includegraphics[width=0.99\hsize]{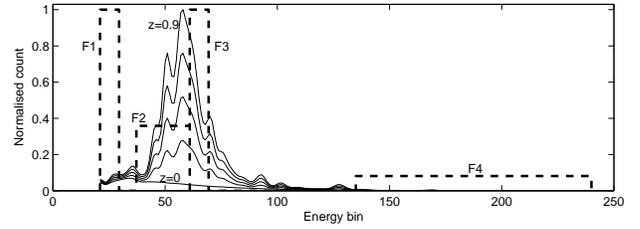} 
\caption{\small A selection of spectra showing variation with metallicity ($Z$=0 to 0.9 solar) with fixed $T$=0.6 keV and fixed $n_H = 0.55\times 10^{25} m^{-2}$. Optimal filter set is shown overlaid.} 
\label{fig4} 
\end{center} 
\end{figure}

\section{Results}

The inversion process outlined in the previous section was applied to a hyper-spectral X-ray image [9], shown in Fig.~3. The raw observation data was reduced using standard techniques to correct for instrumental response, and the background detected from the sky and detector was subtracted using blank field observations. Spectra were extracted in the same energy range as the synthetic spectra, from seven annuli, centred on  the centroid of emission,  of radii 20, 36, 52, 70, 110, 170 and 252 arcsec.

\subsection{Optimal filters}

In order to construct the optimal filters, 1000 synthetic spectra were generated for X-ray emitting optically thin plasmas as described in section 3.1. Parameter space was discretised into a 10x10x10 grid linearly spaced in each parameter to cover the respective parameter ranges ($T$=0.6 to 1.5 keV, $Z$=0.0 to 0.9 and $n_H$=0 to 1.8$\times10^{25} m^{-2}$). Poisson errors, corresponding to a total count of $10^4$ per spectrum, were assumed in order to define the wavelength dependent error associated with the spectral data.

Following the process outlined in \S3.3 and after approximately $1.3\times10^7$ Jacobian constructions, the filter set which both guarantees a non zero Jacobian and minimises the sum (over model parameter space) of parameter recovery errors was found and is shown in Fig.~4 overlaid on a selection of model spectra. The four filters in the set are marked F1 to F4 with F4 used to construct the denominator in quotient space.

\subsection{Parameter recovery}

In order to test the technique, parameter recovery was implemented for the spectra extracted from the seven annuli of the galaxy group HCG~62 (Fig.~3). The spectrum from each region was obtained by integrating the signal over that region. Using the optimal filter set derived in 4.1 the quotient vector for each region was produced and then, using the inverse method (section 3.3), back projected to yield $T$, $Z$ and $n_H$ and their associated errors. The results for the seven annuli are depicted in Figs.~ 5(a), 5(b) and 5(c) along with the  results of the original algorithm for comparison.

\begin{figure}[ht!]
\begin{center}
\includegraphics[scale=0.40]{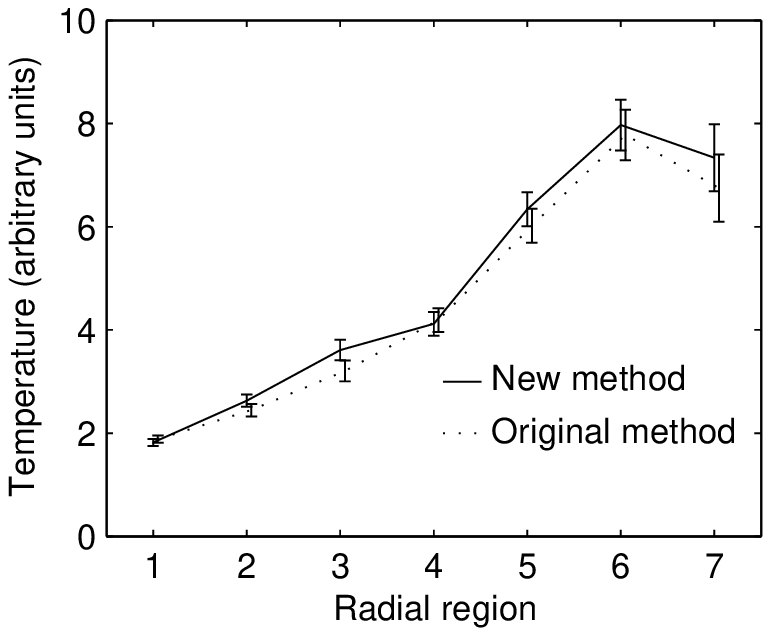}
\includegraphics[scale=0.40]{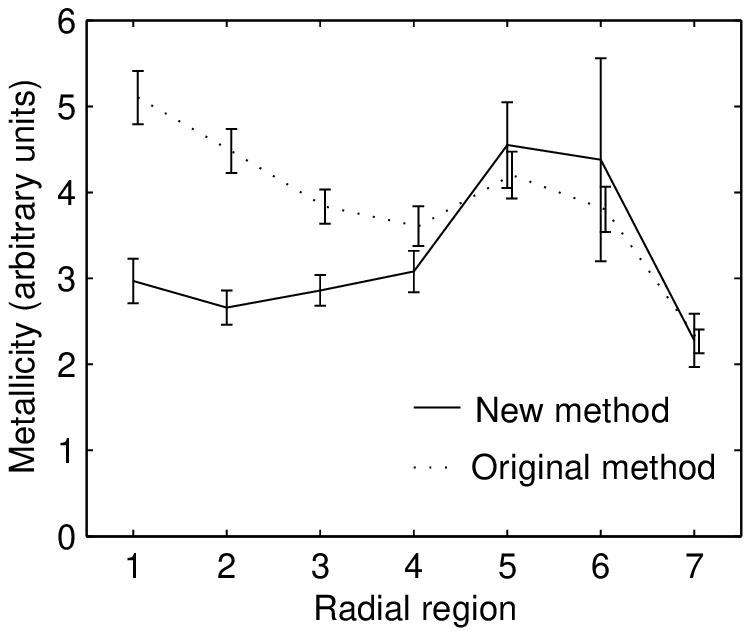}
\includegraphics[scale=0.40]{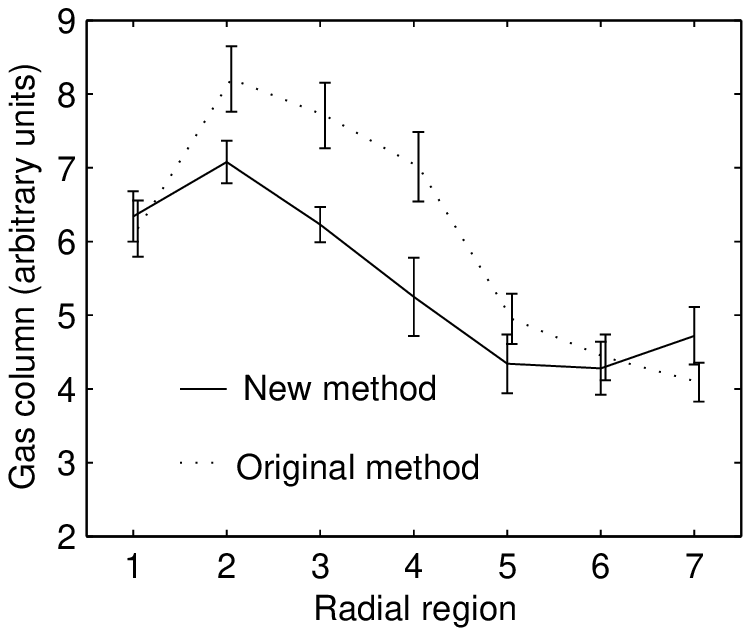}
\label{Fig5} 
\caption{\small Recovered parameter values and associated errors in the seven annular regions for (a) temperature ($T$); (b) metallicity ($Z$); and (c) intervening gas column ($n_H$).}
\end{center}
\end{figure}

\section{Discussion and conclusions}

We have proposed a fast inversion technique which allows most of the essential physical information present in X-ray spectra of hot gas to be presented in the form of parametric maps. This approach improves on the existing technique in two aspects. First, the laborious task of iteratively adjusting the three parameters and a scaling factor to ``best fit" a modelled spectrum to experimental data can be replaced with a simple application of pre-calculated filters to the experimental spectrum and the inverse mapping of the subsequent quotient vector to parameter space. Secondly, since the images can be manipulated using standard image processing techniques before the inversion is performed, our approach permits us to produce stabilised and continuous maps of the physical properties which are important to astrophysicists. This provides a more useful product than current techniques, and involves less computational effort. 

At present the mapping procedure is not guaranteed to produce perfect results (Fig.~5) but the technique was originally designed for applications in visible light where the use of contiguous normalized filters was necessary to enable practical application in digital photography. Such limitations are not relevant to this work and we already envisage splitting filters to make best use of knowledge of the characteristic emission energies of important metallic heavy elements in the hot plasmas of interest. Further improvements in the inverse mapping can be achieved by choosing a variable separation in our parameter space distributions (when constructing the model) so as to more evenly populate quotient space. Both these straightforward modifications will only introduce delays into the one-off filter construction and mapping function definition and will have no impact on the time required to invert experimental spectra.


 The method is generic and has been successfully applied to extract parameters of diagnostic importance from the colour images of the skin [4].

\section*{Acknowledgments}
This work has been supported by the Leverhulme Trust, grant number F00094M.

\end{document}